\documentclass[conference]{IEEEtran}

\newcommand{\subparagraph}{}

\IEEEoverridecommandlockouts
\usepackage{cite}

\usepackage{amsmath,amssymb,amsfonts}
\usepackage{algorithmic}
\usepackage{textcomp}
\usepackage{xcolor}
\def\BibTeX{{\rm B\kern-.05em{\sc i\kern-.025em b}\kern-.08em
    T\kern-.1667em\lower.7ex\hbox{E}\kern-.125emX}}

\usepackage[utf8]{inputenc}
\usepackage{amsthm}
\usepackage{physics}
\usepackage{mathtools}
\usepackage{graphicx}
\usepackage{hyperref}
\usepackage{bbm}
\usepackage{bm}
\usepackage{titlesec}
\usepackage{subcaption}
\usepackage{mathtools}
\usepackage[ruled,vlined]{algorithm2e}

\usepackage[section]{placeins} %[subsection]{placeins} gives an error

\begin{document}

\title{Digital zero noise extrapolation for \\ quantum error mitigation\\
\thanks{U.S. Department of Energy, Office of Science, Office of Advanced Scientific Computing Research, Accelerated Research in Quantum Computing under Award Number DE-SC0020266}
}

\author{
    \IEEEauthorblockN{
    Tudor Giurgica-Tiron\IEEEauthorrefmark{1},
    Yousef Hindy\IEEEauthorrefmark{1},
    Ryan LaRose\IEEEauthorrefmark{2}\IEEEauthorrefmark{4},
    Andrea Mari\IEEEauthorrefmark{2}\IEEEauthorrefmark{5},
    William J. Zeng\IEEEauthorrefmark{1}\IEEEauthorrefmark{2}\IEEEauthorrefmark{3}}
    
    \IEEEauthorblockA{\IEEEauthorrefmark{1}\textit{Stanford University, Palo Alto, CA}}
    
    \IEEEauthorblockA{\IEEEauthorrefmark{2}\textit{Unitary Fund}}
    
    \IEEEauthorblockA{\IEEEauthorrefmark{3}\textit{Goldman, Sachs \& Co, New York, NY}}
    
    \IEEEauthorblockA{\IEEEauthorrefmark{4}\textit{Michigan State University, East Lansing, MI}}
    
    \IEEEauthorblockA{\IEEEauthorrefmark{5}\textit{Xanadu, Toronto, Ontario, Canada}}
}

\maketitle

\begin{abstract}
Zero-noise extrapolation (ZNE) is an increasingly popular technique for mitigating errors in noisy quantum computations without using additional quantum resources. We review the fundamentals of ZNE and propose several improvements to noise scaling and extrapolation, the two key components in the technique. We introduce unitary folding and parameterized noise scaling. These are digital noise scaling frameworks, i.e. one can apply them using only gate-level access common to most quantum instruction sets.
We also study different extrapolation methods, including a new adaptive protocol that uses a statistical inference framework. Benchmarks of our techniques show error reductions of 18X to 24X over non-mitigated circuits and demonstrate ZNE's effectiveness at larger qubit numbers than have been tested previously.
In addition to presenting new results, this work is a self-contained introduction to the practical use of ZNE by quantum programmers.
\end{abstract}


\section{Introduction}

As quantum hardware becomes available in the noisy intermediate-scale quantum computing (NISQ) era~\cite{Preskill2018QuantumBeyond}, it is inevitable that today's quantum programmer must deal with errors. In the long run, fault-tolerance and quantum error-correction have the potential to arbitrarily reduce logical errors~\cite{Terhal2015QuantumMemories, Gottesman2010AnComputation, Fowler2012SurfaceComputation}. However, a scalable logical qubit has yet to be demonstrated. Thus the savvy quantum programmer should make use of \emph{error-mitigating} techniques that give practical benefits, even if they do not arbitrarily suppress errors in the asymptotic limit. In the NISQ era, every constant factor counts.

There are many examples of error-mitigating techniques, including probabilistic error cancellation~\cite{Temme2017ErrorCircuits, endo2018practical}, randomized compiling~\cite{wallman2016noise}, Pauli-frame randomization~\cite{knill2005quantum}, dynamical decoupling~\cite{santos2005dynamical, viola2005random, pokharel2018demonstration, sekatski2016dynamical}, quantum optimal 
control~\cite{ball2020software, green2013arbitrary}, etc. In this work, we focus on the specific error-mitigating technique known as {\it zero-noise extrapolation}.

Zero-noise extrapolation (ZNE) was introduced concurrently in~\cite{Temme2017ErrorCircuits} and~\cite{Li2017EfficientMinimization}. In ZNE, a quantum program is altered to run at different effective levels of processor noise. The result of the computation is then extrapolated to an estimated value at a noiseless level. More formally, one can parameterize the noise-level of a quantum system with a dimensionless 
 scale factor $\lambda$. For $\lambda=0$ the noise is removed, while for $\lambda=1$ the true noise-level of the physical hardware is matched. For example,
$\lambda$ could be a multiplicative factor that scales the dissipative terms of a master equation \cite{Temme2017ErrorCircuits}. More generally, $\lambda$ could represent a re-scaling of any physical quantity which introduces some noise in the quantum computation: the calibration uncertainty of variational parameters, the temperature of the quantum processor, etc.

\begin{figure}[t]
    \centering
    \includegraphics[width=0.8 \columnwidth]{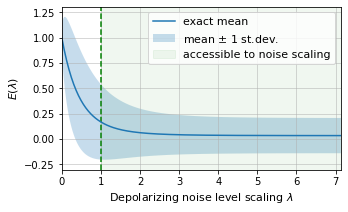}
    \caption{An example of the change of an expectation value, $E(\lambda)$, with the underlying scaling $\lambda$ of the depolarizing noise level. Here the simulated base noise value is 5\% (marked by the green dashed vertical line). ZNE increases that noise and back extrapolates to the $\lambda=0$ expectation value. In this example, an accurate extrapolation should be non-linear and take advantage of a known asymptotic behavior.}
    \label{fig:example}
\end{figure}

For a given quantum program, we can measure an arbitrary expectation value $E(\lambda)$. By construction, $E(1)$ represents the expectation value evaluated with the natural noise of the hardware, whereas $E(0)$ denotes the noiseless observable which, despite being not directly measurable, we would like to estimate.

To implement ZNE, one needs a direct or indirect way to scale the quantum computation's noise level to values of $\lambda$ larger than one. With such a method, ZNE can be implemented in two main steps:

\begin{enumerate}
    \item {\bf Noise-scaling:} Measure $E(\lambda)$ at $m$ different values of $\lambda\ge 1$.
    \item {\bf Extrapolation:} Infer $E(0)$ from the $m$ expectation values measured in previous step. 
\end{enumerate}

Figure~\ref{fig:example} shows an example \emph{noise curve} given by scaling depolarizing noise for a randomized benchmarking circuit.

In this work, we introduce improvements to both noise-scaling and extrapolation methods for quantum error mitigation. In Section~\ref{sec:unitary-folding} we introduce unitary folding, a framework for digital noise scaling of generic gate noise. In Section~\ref{sec:parameter-noise-scaling} we introduce a general method for ZNE tailored for a specific kind of errors: calibration noise. We then move to the extrapolation step of ZNE, which we characterize as an inference problem. We study non-adaptive (Section~\ref{sec:non-adaptive}) extrapolation methods and introduce adaptive (Section~\ref{sec:adaptive}) extrapolation to improve performance and reduce resource overhead for ZNE.

\section{Noise Scaling Methods}
\label{sec:noise-scaling}

In~\cite{Temme2017ErrorCircuits} and~\cite{ Kandala2019ErrorProcessor} a time-scaling approach implements the scaling of effective noise on the back-end quantum processor. Control pulses for each gate are re-calibrated to execute the same unitary evolution but applied over a longer amount of time. This effectively scales up the noise.
While successfully used to suppress errors in single and two-qubit quantum programs on a superconducting quantum processor~\cite{Kandala2019ErrorProcessor}, time-scaling has some disadvantages:
\begin{itemize}
    \item It requires programmer access to low-level physical-control parameters. This level of access is not available on all quantum hardware and breaks the gate model abstraction.
    \item Control pulses must be re-calibrated for each time duration and error-scaling. This calibration can be resource intensive.
\end{itemize}

Instead, we study alternative approaches that require only a gate-level access to the system. Rather than increasing the time duration of each gate, we increase the total number of gates or, similarly, the circuit depth. This procedure is similar to what is usually done by a {\it quantum compiler} but with the opposite goal: instead of optimizing a circuit to reduce its depth or its gate count, we are interested in ``de-optimizing'' to increase the effect of noise and decoherence. We use the term~\emph{digital} to describe noise-scaling techniques that manipulate just the quantum program at the instruction set layer. Their advantage is that they can be used with the gate model access that is common to most quantum assembly languages~\cite{Smith2016AArchitecture, McKay2017EfficientComputing, fu2019eqasm}. Low level access to pulse shaping and detailed physical knowledge of quantum processor physics is no longer required. Our digital framework incorporates and generalizes some recent related work~\cite{he2020resource, dumitrescu2018cloud}.

\subsection{Unitary Folding}
\label{sec:unitary-folding}
We describe two methods--circuit folding and gate folding--for scaling the effective noise of a quantum computation based on {\it unitary folding}, i.e., replacing a unitary circuit (or gate) $U$ by:
\begin{equation}\label{Ufold}
U \rightarrow U(U^\dag U)^n,
\end{equation}
where $n$ is a positive integer. In an ideal circuit, since $U^\dag U$ is equal to the identity, this folding operation has no logical effect. However, on a real quantum computer, we expect that the noise increases since the number of physical operations scales by a factor of $1 + 2 n$. This effect is clearly visible in the quantum computing experiment reported in Figure \ref{fig:ibmq}.

A similar trick was used in Ref. \cite{he2020resource, dumitrescu2018cloud}, where noise was artificially increased by inserting pairs of CNOT gates into quantum circuits. In our framework, $U$ can represent the full input circuit or, alternately, some local gates which are inserted with different strategies.

\subsubsection{Circuit folding}

\label{sec:circuit-folding}
Assume that the circuit is composed of $d$ unitary layers:
\begin{equation}\label{circuit_layers}
U = L_d ... L_2 L_1, 
\end{equation}
where $d$ represents the {\it depth} of the circuit and each block $L_j$ can either represent a single layer of operations or just a single gate.

In circuit folding, the substitution rule in Eq.~\eqref{Ufold} is applied globally, i.e., to the entire circuit. This scales the effective depth by odd integers. In order to have a more fine-grained resolution of the scaling factor, we can also allow for a final folding applied to a subset of the circuit corresponding to its last $s$ layers. The general {\it circuit folding} replacement rule is therefore:

\begin{equation}\label{circuit_folding}
U \rightarrow U(U^\dag U)^n L_d^\dag L^\dag_{d-1} \dots L_s^\dag L_s \dots L_{d-1} L_d.
\end{equation}
The total number of layers of the new circuit is $d(2n + 1) + 2 s$. This means that we can stretch the depth of a circuit up to a scale resolution of $2/d$, i.e., we can apply the scaling $d \rightarrow \lambda d$, where:
\begin{equation}\label{lambda_circuit_folding}
\lambda = 1 + \frac{2 k}{d}, \quad k=1,2,3,  \dots .
\end{equation}
Conversely, for every real $\lambda$, one can apply the following procedure:

\begin{enumerate}
\item Determine the closest integer $k$ to the real quantity $ d (\lambda - 1) / 2 $.
\item Perform an integer division of $k$ by $d$. The quotient corresponds to $n$, while the reminder to $s$.
\item Apply $n$ integer foldings and a final partial folding as described in Eq.~\eqref{circuit_folding}.
\end{enumerate}

From a physical point of view, the circuit folding method corresponds to repeatedly driving the Hamiltonian of the qubits forwards and backwards in time, such that the ideal unitary part of the dynamics is not changed while the non-unitary effect of the noise is amplified.

\subsubsection{Gate (or Layer) folding}

Instead of globally folding a quantum circuit, appending the folds at the end, one could fold a subset of individual gates (or layers) in place.
Let us consider the circuit decomposition of Eq.~\eqref{circuit_layers} where we can assume that each unitary operator $L_j$ represents just a single gate applied to one or two qubits of the system or, alternatively, each $L_j$ could be a layer of several gates.

If we apply the replacement rule given in Eq.~\eqref{Ufold} to each gate (or layer) $L_j$ of the circuit, it is clear that the initial number of gates (layers) $d$ is scaled by an odd integer $1 + 2n$. Similarly to the case of circuit folding, we can add a final partial folding operation to get a scaling factor which is more fine grained. In order to achieve such ``partial'' folding, let us define an arbitrary subset $S$ of the full set of  indices $\{1, 2, \dots d\}$, such that its number of elements is a given integer $s = |S|$.
In this setting, we can define the following {\it gate (layer) folding} rule:
\begin{equation}\label{gate_folding}
\forall j \in \{1,2, \dots d\}, \quad L_j \rightarrow 
\left\{ 
\begin{array}{l l}
L_j(L_j^\dag L_j)^n   &{\rm if} j \notin S, \\
\\
L_j(L_j^\dag L_j)^{n+1}  &{\rm if} j  \in  S. \\
\end{array}
\right. 
\end{equation}

Depending on how we chose the elements of the subset $S$, different noise channels will be added at different positions along the circuit  and so we can have different results. The optimal choice may depend on the particular circuit and noise model. We focus on three different ways of selecting the subset of gates (layers) to be folded: {\it from left}, {\it from right} and {\it at random}. Depending on the method, the prescription for selecting the subset $S$ of indices is reported in Table~\ref{tab:gate_folding_methods}. 

\begin{table}[t]
\caption{Different methods for implementing gate (or layer) folding}
\centering
\begin{tabular}{l|l}
%\cline{1-3}
\bf Method       & \bf Subset of indices to fold \\ %\cline{1-3}
\hline
\hline
From left 	& $S=\{1, 2, \dots, s\}$ 	\\
From right 	& $S=\{d, d-1, \dots, d - s +1\}$ 	\\
At random 	& $S=$ $s$ different indices randomly sampled \\
			& without replacement from $\{1, 2, \dots, d\}$.	
\end{tabular}

\label{tab:gate_folding_methods}
\end{table}

It is easy to check that the number of gates (or layers), obtained after the 
application of the gate folding rule given in Eq.~\eqref{gate_folding} is $d (2n + 1) + 2s$. This is exactly the same number obtained after the application of the global circuit-folding rule given in Eq.~\eqref{circuit_folding}. As a consequence, the number of gates (layers) is still stretched by a factor $\lambda$, i.e., $d \rightarrow \lambda d$, where
$\lambda$ can take the specific values reported in Eq.~\eqref{lambda_circuit_folding}. Moreover, if we are given an arbitrary $\lambda$ and we want to determine the values of $n$ and $s$, we can simply apply the same procedure that was given in the case of circuit-folding.

While preparing this manuscript we became aware of~\cite{he2020resource} whose technique is similar to our gate folding (at random). The main difference is that~\cite{he2020resource} focuses mainly on CNOT gates and uses random sampling with replacement, in our case any gate (or layer) can be folded and the sampling is performed without replacement. The rationale of this choice is to sample in a more uniform way the input circuit, and to converge smoothly to the odd integer values of $\lambda= 1+2n$ where all the input gates are folded exactly $n$ times.

\subsubsection{Advantages and limitations of unitary folding}

The main advantage of the unitary folding approach is that is is digital, i.e., noise is scaled using a high level of abstraction from the physical hardware. Moreover, it can be applied without knowing the details of the underlying noise-model. It is natural to ask: how justified is this approach physically? Does unitary folding actually correspond to an effective scaling of the physical noise of the hardware?

For example, unitary folding may fail to amplify systematic and coherent errors since applying the inverse of a gate will usually  {\it undo} such errors instead of increasing them.  It is also clear that unitary folding is not appropriate to scale state preparation and measurement (SPAM) noise, since this noise is independent of the circuit depth. Instead, we expect that unitary folding can be used for scaling incoherent noise models which are associated both to the application of individual gates and/or to the time-length of the overall computation. The more we increase the depth of the circuit, the more such kinds of noise are usually amplified. In this work this intuition is confirmed by numerical and experimental examples in which unitary folding is successfully used for implementing ZNE (see Figures \ref{fig:rb}, \ref{fig:random_6qubit_circuits_histogram}, \ref{fig:maxcut-violin} and
 \ref{fig:ibmq}).

The effect of unitary folding can be analytically derived when the noise-model for each gate $L_j$ is a global depolarizing channel with a gate-dependent parameter $p_j \in [0, 1]$, acting as:
\begin{equation}
\label{eq:depolarizing}
\rho \xrightarrow{\text{noisy gate}}  p_j L_j \rho L_j^\dag + (1-p_j) \mathbb{I}/D,
\end{equation}
where $D$ is the dimension of the Hilbert space associated to all the qubits of the circuit.  Since the depolarizing channel commutes with unitary operations, we can postpone the noise channels of all the gates until the end of the full circuit $U$, resulting into a single final depolarizing channel:
\begin{equation}
\rho \xrightarrow{\text{noisy circuit}}  p U \rho U^\dag + (1-p) \mathbb{I}/D,
\end{equation}
where $p = \Pi_j p_j$ is the product of all the gate-dependent noise parameters $p_j$. This simple 
commutation property does not hold for local depolarizing noise, unless we are dealing with singe-qubit circuits. 

Consider what happens if we apply unitary folding with a scale factor $\lambda = 1 + 2n$ (odd positive integer). For both the circuit folding and the gate folding methods, defined in Eq.~\eqref{circuit_folding} and 
\eqref{gate_folding} respectively, the final result is exactly equivalent to an exponential scaling of all the depolarizing parameters of each gate $p_j \rightarrow p_j^\lambda$ or, equivalently, to the global operation:
\begin{equation}\label{uf_depo_state}
\rho \xrightarrow{\text{noise + unitary folding}}  p^\lambda U \rho U^\dag + (1-p^\lambda) \mathbb{I}/D.
\end{equation}
This implies that unitary folding is equivalent to an exponential parameterization of the noise level $p$, and so any expectation value is also scaled according to an exponential ansatz:
\begin{equation}\label{uf_depo_exp}
E(\lambda) = a + b p^\lambda,
\end{equation}
which we can fit and extrapolate according to the methods discussed in the Sections \ref{sec:non-adaptive} and \ref{sec:adaptive}.

Equations~\eqref{uf_depo_state} and \eqref{uf_depo_exp} are valid only for depolarizing noise and for odd scaling factors $\lambda$. For gate-independent depolarizing noise, the global parameter $p$ is a function of the total number of gates only. This means that all the folding methods (circuit, from left, from right and at random) become equivalent, and induce the exponential scalings of Eqs.~\eqref{uf_depo_state} and \eqref{uf_depo_exp})
for all values of $\lambda$.

\subsubsection{Numerical Results}

\begin{figure}[t]
    \centering
    \includegraphics[width=0.9 \columnwidth]{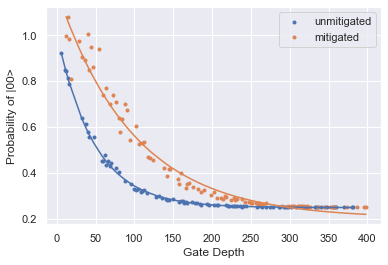}
    \caption{A comparison of two qubit randomized benchmarking with and without error-mitigation. Data is taken by density matrix simulation with a 1\% depolarizing noise model. The unmitigated simulation results in a randomized benchmarking decay of $97.9\%$. Mitigation is applied using circuit folding and an order-2 polynomial extrapolation at $\lambda=1, 1.5, 2.0$. With mitigation the randomized benchmarking decay improves to $99.0\%$. Since we do not impose any constraint on the  domain of the extrapolated results, some of the mitigated expectation values are slightly beyond the physical upper limit of $1$. This is an expected effect of the noise introduced by the extrapolation fit. If necessary, one could enforce the result to be physical by using a more advanced Bayesian estimator.}
    \label{fig:rb}
% Data for this figure was generated by
% https://github.com/unitaryfund/mitiq-examples/blob/master/random/RB_data_v2.ipynb
\end{figure}

We executed density matrix simulations using unitary folding for zero-noise extrapolation. Broadly these results show that unitary folding is effective in a variety of situations. Furthermore, we benchmark on both random circuits and a variational algorithm at 6 and more qubits. This extends previous work that focuses on the single and two qubit cases~\cite{Temme2017ErrorCircuits, Li2017EfficientMinimization, Kandala2019ErrorProcessor, endo2018practical}. Figure~\ref{fig:rb} shows a simulated two qubit randomized benchmarking experiment under 1\% depolarizing noise with and without error-mitigation. Noise was scaled using circuit folding as described in Section~\ref{sec:circuit-folding}.

\begin{figure}[tb]
    \centering
    \includegraphics[width=0.9 \columnwidth]{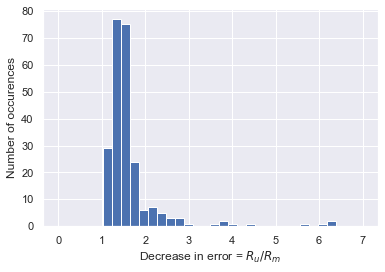}
    \caption{A comparison of improvements from ZNE (using quadratic extrapolation with folding from left) averaged across all output bitstrings from 250 random six-qubit circuits. Results are from exact density matrix simulations with a base of 1\% depolarizing noise. The horizontal axis shows a ratio of $L_2$ distances from the noiseless probability distribution and the vertical axis shows the frequency of obtaining this result. ZNE improves on the noisy result by factors of 1-7X. The average mitigated error is $0.075\pm0.035$, while the unmitigated errors average $0.114\pm0.050$. Each circuit has 40 moments with single-qubit gates sampled randomly from $\{H, X, Y, Z, S, T\}$ and two-qubit gates sampled randomly from $\{\text{iSWAP}, \text{CZ}\}$ with arbitrary connectivity.}
    \label{fig:random_6qubit_circuits_histogram}
\end{figure}

Figure~\ref{fig:random_6qubit_circuits_histogram} shows the distribution of noise reduction by ZNE with circuit folding on randomly generated six qubit circuits. Let $E_m$ be the mitigated expectation value of a circuit after zero-noise extrapolation. Then $R_m = |E_m-E(0)|$ is the absolute value of the error in the mitigated expectation and $R_u = |E(1)-E(0)|$ is the absolute value of the error of the unmitigated circuit. The improvement from ZNE is quantified as $R_u/R_m$.

Table~\ref{tab:benchmarks} (see Section~\ref{sec:non-adaptive}) provides a comparison different combinations of folding and extrapolation techniques on a set of randomized benchmarking circuits.

% Data for this figure was generated by
% https://github.com/unitaryfund/mitiq-examples/blob/master/random/random_circ_2.py
% Data was plotted with https://github.com/unitaryfund/mitiq-examples/blob/master/random/RB_plots.ipynb

\begin{figure}[tb]
    \centering
    \includegraphics[width=0.9 \columnwidth]{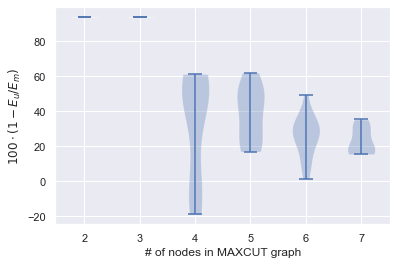}
    \caption{Percent closer to optimal on random MAXCUT executions. 14 Erdos-Renyi random graphs were generated at each number $n$. Each random graph has $n$ nodes and $n$ edges. QAOA was then run (with $p=2$ QAOA steps) and optimized using Nelder-Mead with and without error mitigation. Results are from exact density matrix simulations with a base of 2\% depolarizing noise. For the mitigated case, we used zero noise extrapolation with global unitary folding for scaling and linear extrapolation at noise scalings of 1, 1.5 and 2. The y axis shows the percent closer to the optimal solution that was gained by ZNE. Here $E_u$ is the absolute error in the unmitigated expectation and $E_m$ is the absolute error in the mitigated expectation. The violin plot shows the distribution of percentage improvements over the 14 sampled instances. Variance is zero for 2 and 3 nodes graphs as there is only a single valid graph with $n$ nodes and edges for $n=2,3$.}
    \label{fig:maxcut-violin}
\end{figure}

Figure~\ref{fig:maxcut-violin} shows the performance of unitary folding ZNE on a variational algorithm. Using exact density matrix simulation we study the percentage closer to optimal achieved by the quantum approximation optimization algorithm~\cite{Farhi2014AAlgorithm} on random instances of MAXCUT. 

% p=2 data generated with https://github.com/unitaryfund/mitiq-examples/blob/master/maxcut/maxcut_benchmark_data_p2.ipynb
% plots made with https://github.com/unitaryfund/mitiq-examples/blob/master/maxcut/maxcut_plots.ipynb 

\subsection{Parameter Noise Scaling}
\label{sec:parameter-noise-scaling}

While unitary folding applies to general classes of noise models, it is reasonable to ask if we can exploit the specific structure of particular noise models. We give an example of this approach by mitigating errors in the stochastic calibration of parametric quantum gates. The error model that we consider is a generalized form of the ``pulse-area'' error, which describes what happens when the physical pulses that generate a particular gate in a quantum processor are slightly mis-calibrated \cite{Barnes2017PulseArea, Mendl2009Unital, Merrill2012Pulses, Wang2012Pulses, Wang2011DynamicalPulses}. This noise could be due to fluctuations in control electronics, or uncertainty about underlying physical parameters (such as qubit frequencies) in available hardware. Furthermore, our model applies also to variational quantum circuits, in which the parametric dependence of the gates is critically accessed by quantum programmers.

In order to apply ZNE in this setting, we need a method for scaling this particular kind of noise source. Instead of changing the structure of the quantum circuit, as in the unitary folding method, we directly inject classical noise into the control parameters. This artificial noise can increase the native noise of the hardware to larger levels, such that ZNE becomes applicable. We call this approach~\emph{parameter noise scaling}.

\subsubsection{Parameter Noise Scaling Theory}
\label{sec:parameter-noise-scaling-theory}

We assume that a quantum gate is parameterized by $l$ classical control parameters $\bm \theta=(\theta_1, \theta_2, \dots \theta_l)$, such that

\begin{align} \label{eq:parametric_gate}
    G(\bm \theta) &= \exp (-i \sum_{j=1}^l \theta_j H_j),
\end{align}
where $H_1, H_2, \dots H_l$ are Hermitian operators. In practice, the parameters $\bm \theta$ can represent the classical controls that the quantum processor needs to tune in order to implement a particular gate. Alternately, $G(\bm \theta)$ could also model a variational gate which can be programmed by the user.

In both cases, it is reasonable to assume that the control parameters can be applied only up to some finite
precision, i.e., that what is actually implemented on the physical system is the gate $G(\bm \theta')$, where

\begin{align} \label{eq:noisy_theta}
    \theta_j'= \theta_j + \hat \epsilon_j,
\end{align}
and $\hat \epsilon_j$ is a random variable with zero-mean and variance $\sigma_j^2$, which represents a stochastic calibration error (note that this is not a constant systematic error). Going forward, we will assume that $\hat \epsilon$ is Gaussian distributed, however the analysis could be generalized to other cases.

Consider the case in which the variances $\sigma_j^2$ associated to all control parameters are known. These variances  could be estimated by performing tomography on repeated applications of the same gate and inferring the distributions of the control parameters. With $\sigma_j^2$ in hand, noise scaling can be directly applied by shifting the control parameters with some additional classical noise $\hat \delta_j$:

\begin{align} \label{eq:delta_addition}
\theta_j'  \xrightarrow{\text{parameter noise scaling}}  \theta_j' + \hat \delta_j
\end{align}
where the $\hat \delta_j$ is sampled from a zero-mean Gaussian distribution with variance $(\lambda - 1) \sigma_j^2$,
such that the variance of the overall noise is scaled by a factor of $\lambda \ge 1$ . Equivalently, the effect
of parameter noise scaling is that of transforming Eq.~\eqref{eq:noisy_theta} into
\begin{align} \label{eq:lambda_theta}
\theta_j'= \theta_j + \sqrt{\lambda} \hat \epsilon_j.
\end{align}

This gives a simple noise scaling procedure for ZNE that can be done without knowing the particular structure of the Hermitian operators $H_j$ and also without knowing the Kraus operators of the corresponding error channel.

However, if we are interested in a density matrix simulation of the quantum circuit, it may still be useful to derive the analytical Kraus operators corresponding to the noise model of Eq.~\eqref{eq:noisy_theta}.
Since in general the operators $H_j$ do not commute with each other, this is a subtle task. 
For simplicity, here we derive analytically the noise channel in the case of a single-parameter {\it rotation-like} gate, i.e., such that it can be expressed as:

\begin{align}\label{eq:rot_like}
G(\theta) = \exp(-i\theta H/2) = \cos(\theta/2) \mathbb{I} - i \sin (\theta/2)H.
\end{align}
This property holds whenever $H^2 = \mathbb{I}$, including the important cases of Pauli or controlled-Pauli rotations.
Moreover, since we are dealing with a single parameter, we can easily factorize the noisy operation as the ideal gate followed by a purely random rotation:

\begin{align}
G(\theta') = G(\hat \epsilon) G(\theta).
\end{align}
We are interested in the effect of the final noisy gate $G(\hat \epsilon)$ on the density matrix of the system. This is given by averaging over the Gaussian probability distribution $p(\epsilon)$ associated to the
random variable $\hat \epsilon$:
\begin{align}
    &\mathcal{E}(\rho)  = \int_{-\infty}^{\infty}  p(\epsilon) G(\epsilon) \rho G^\dagger(\epsilon) d\epsilon   \nonumber \\
    & =   \int_{-\infty}^{\infty} p(\epsilon) \left[\cos^2 (\epsilon) \rho + i \sin(2\epsilon) [\rho, H] + \sin^2(\epsilon) H \rho H \right] d\epsilon \nonumber \\
    & = (1-Q)\rho + Q H \rho H, \label{eq:kraus_q}
\end{align}
where $Q = \frac{1}{2}(1 - e^{-2\sigma^2})$ is a simple function of the noise variance $\sigma^2$. For a single parameter and for rotation-like gates, the effect of the noise-model defined in Eq.~\eqref{eq:noisy_theta} is a 
quantum channel with only two Kraus operators. The channel is probabilistic mixture of the identity operation (with probability $1-Q$) and the unitary $H$ (with probability $Q$). In the limit of a small noise variance $\sigma^2$, even if the gate does not obey the rotation-like property \eqref{eq:rot_like}, the quantum channel is still correctly approximated by Eq.~\eqref{eq:kraus_q} up to $O[(\sigma^2)^2]$ corrections. The same derivation can be applied also for different probability distributions of the noise. 

\begin{figure}[bt]
    \centering
    \includegraphics[width=0.9 \columnwidth]{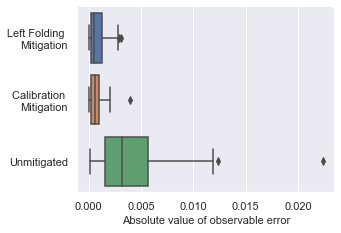}
    \caption{Errors without mitigation, with parameter noise mitigation, and with unitary folding mitigation. Each distribution is over 50 random six-qubit circuits with random computational basis observables mitigated using gate folding from left and parameter noise scaling. The underlying noise is an angle noise channel at $\sigma^2=0.001$. We used 
a linear extrapolation with noise scale factors $\lambda=\{1,2,3\}$. Results were obtained with exact density matrix simulations and are presented in box plots of the distribution across the random circuits. The diamond points are outliers.}
    \label{fig:angle-on-angle}
 \end{figure}

Figure~\ref{fig:angle-on-angle} uses exact density matrix simulation to estimate the performance of calibration noise scaling. Here we plot the absolute value of observable error ($|E_m-E(0)|$) for randomly generated six qubit circuits. We see that calibration mitigation performs as well as unitary folding mitigation but it has the
advantage of not adding new gates to the circuit. Thus it is likely to be less sensitive to other sources of noise such as decoherence.

% data taking and plotting: % https://github.com/unitaryfund/mitiq-examples/tree/master/calibration

\section{Non-adaptive extrapolation methods: Zero noise extrapolation as statistical inference}
\label{sec:non-adaptive}

\begin{algorithm}[t]
 \KwData{A set of increasing noise scale factors $\bm \lambda = \{\lambda_1, \lambda_2, \dots \lambda_m \}$, with $\lambda_j \ge 1$ and fixed number of samples $N$ for each $\lambda_j$.}
 \KwResult{A mitigated expectation value}
 $\bm y \longleftarrow \emptyset$\;
 \Begin{
 \For{$\lambda_j \in \bm \lambda$}{
    $y_j \longleftarrow ComputeExpectation(\lambda_j, N)$\;
    $Append\left(\bm y, y_j\right)$\;
    }
 \tcc{Abitrary best fit algorithm (e.g., least squares)}
 $\bm \Gamma^*\longleftarrow BestFit(E_{\rm model}(\lambda; {\bm \Gamma}),  (\bm \lambda, \bm y))$\;
 \KwRet{$E_{\rm model}(0; {\bm \Gamma^*})$}\;
 }

\caption{Generic non-adaptive extrapolation}
\label{alg:non-adaptive}
\end{algorithm}

In Section~\ref{sec:noise-scaling}, we discussed several methods to scale noise. In this section we study, from an estimation theory perspective, the second component of ZNE: extrapolating the measured data to the zero-nose limit.

We assume that the output of the quantum computation is a single expectation value $E(\lambda)$, where $\lambda$ is the noise scale factor. This expectation could be the result of a single quantum circuit or some combinations of quantum circuits with classical post-processing.
The expectation value $E(\lambda)$ is a real number which, in principle, can only be estimated in the limit of infinite measurement samples. In a real situation with $N$ samples, only a statistical estimation of the expectation value is actually possible:
\begin{align}\label{estim_E}
\hat E(\lambda) = E(\lambda) + \hat \delta,
\end{align}
where $\hat \delta$ is a random variable with zero mean and variance $\sigma^2 = \mathbbm{E}(\hat \delta^2)=\sigma_0^2/N$, with $\sigma_0^2$ corresponding to the single-shot variance. In other words, we can sample a real prediction $y$ from the probability distribution:
\begin{align}\label{estim_dist}
P(\hat E(\lambda) = y) = \mathcal N (E(\lambda) - y, \sigma^2 ),
\end{align}
where $\mathcal N(\mu, \sigma^2)$ is a generic distribution (typically Gaussian), with mean $\mu$ and variance $\sigma^2 = \sigma_0^2/N$.

Given a set of $m$ scaling parameters $\bm \lambda = \{\lambda_1, \lambda_2, \dots \lambda_m \}$, with $\lambda_j \ge 1$, and the corresponding results 
\begin{align}
    {\bf y} = \{y_1, y_2, \dots\ y_m \},
\end{align}
the ZNE problem is to build a good estimator $\hat E(0)$ for $E(\lambda = 0)$, such that its bias 
\begin{align}\label{bias_zne}
{\rm Bias}(\hat E(0)) = \mathbbm{E} (\hat E(0) -  E(0)),
\end{align}
and its variance
\begin{align}\label{var_zne}
{\rm Var} (\hat E(0))= \mathbbm{E} (\hat E(0)^2) - \mathbbm{E}(\hat E(0))^2,
\end{align}
are both reasonably small. More precisely, a typical figure of merit for the  quality the estimator is its mean squared error with respect to the true unknown parameter:
\begin{align}\label{mse_zne}
{\rm MSE}(\hat E(0)) &= \mathbbm{E} (\hat E(0) - E(0))^2 \\
&= {\rm Var} (\hat E(0)) + {\rm Bias}(\hat E(0))^2.
\end{align}

If the expectation value $E(\lambda)$ can be an arbitrary function of $\lambda$ without any regularity assumption, then zero-noise extrapolation is impossible. Indeed its value at $\lambda=0$ would be arbitrary and unrelated to its values at $\lambda \ge 1$.
However from physical considerations, it is reasonable to have a model for $E(\lambda)$, e.g., we can assume a linear, a polynomial or an exponential dependence with respect to $\lambda$. For example, for a depolarizing noise model, one can use the exponential ansatz given in Eq.~\eqref{uf_depo_exp}.

If we chose a generic model $E_{\rm model}(\lambda; {\bf \Gamma})$ for the quantum expectation value, where $\bf \Gamma$ represents the model parameters, then the zero-noise-extrapolation problem reduces to a regression problem. Algorithm~\ref{alg:non-adaptive} is the general form for a non-adaptive ZNE. Alternatively, the scale factors $\lambda_j$ and the associated numbers of samples $N_j$ can be chosen in an adaptive way, depending on the results of intermediate steps. This adaptive extrapolation method is studied in more details in Section \ref{sec:adaptive}.

We focus on two main non-adaptive models, the polynomial ansatz and the poly-exponential ansatz. These two general models, give rise to a large variety of specific extrapolation algorithms. Some well known methods, such as Richardson's extrapolation, are particular cases. Some other  methods have, to our knowledge, not been applied before for quantum error mitigation.

\subsection{Polynomial extrapolation}
The polynomial extrapolation method is based on the following polynomial model of degree $d$: 
\begin{align}\label{poly_ansatz}
E_{\rm poly}^{(d)}(\lambda) = c_0 + c_1 \lambda + \dots c_d \lambda^d,
\end{align}
where $c_0, c_1, \dots c_d$ are $d + 1$ unknown real parameters. This essentially corresponds to a Taylor series approximation and is physically justified in the weak noise regime.

In general, the problem is well defined only if the number of data points $m$ is at least equal to the number of free parameters $d + 1$. As opposed to Richardson's extrapolation \cite{Temme2017ErrorCircuits}, a useful feature of this method is that we can keep the extrapolation order $d$ small but still use a large number of data points $m$. This avoids an over-fitting effect: if we increase the order $d$ by too much, then the model is forced to follow the random statistical fluctuations of our data at the price of a large generalization error for the zero-noise extrapolation.
In terms of the inference error given in Eq.~\eqref{mse_zne}, if we increase $d$ by too much, then the bias is reduced but the variance can grow so much that the total mean squared error is actually increased. 

\subsection{Linear extrapolation}

Linear extrapolation is perhaps the simplest method and is a particular case of polynomial extrapolation. It corresponds to the model:
\begin{align}\label{linear_ansatz}
E_{\rm linear}(\lambda) = E_{\rm poly}^{(d=1)}(\lambda) = c_0 + c_1 \lambda.
\end{align}
In this case a simple analytic solution exists, corresponding to the ordinary least squared estimator of the intercept parameter:
\begin{align}\label{linear_analytic}
\hat E_{\rm linear}(0) &=  \bar y - \frac{S_{\lambda y}}{S_{\lambda \lambda}} \bar x,
\end{align}
where
\begin{align}\label{ols}
\bar \lambda &= \frac{1}{m}\sum_j \lambda_j, 
&\bar y &= \frac{1}{m}\sum_j y_j, \nonumber \\
S_{\lambda y} &= \sum_j (\lambda_j - \bar \lambda)(y_j - \bar y),
&S_{\lambda \lambda} &= \sum_j (\lambda_j - \bar \lambda)^2.
\end{align}

With respect to the zero noise value of the model $E_{\rm linear}(0)$, the estimator is unbiased. If the statistical uncertainty $\sigma^2$ for each $y_j$ is the same, the variance for $\hat E_{\rm linear}(0)$ is: 
 
\begin{align}\label{var_analytic}
{\rm Var}[\hat E_{\rm linear}(0)] &= \sigma^2 \left[ \frac{1}{m} + \frac{\bar \lambda^2}{S_{\lambda \lambda}} \right].
\end{align}

\subsection{Richardson extrapolation}
\label{sec:richardson}
Richardson's extrapolation is also a particular case of polynomial extrapolation where $d=m-1$, i.e., the order is maximized given the number of data points:
\begin{align}\label{richardson_ansatz}
E_{\rm Rich}(\lambda) =E_{\rm poly}^{(d=m-1)}(\lambda)= c_0 + c_1 \lambda + \dots c_{m-1} \lambda^{m-1}.
\end{align}
This is the only case in which the fitted polynomial perfectly interpolates the $m$ data points such that, in the ideal limit of an infinite number of samples $N \rightarrow \infty$, the error with respect to the true expectation value is by construction $O(m)$.
Using the interpolating {\it Lagrange polynomial}, the estimator can be explicitly expressed as:
\begin{align}\label{richardson_analytic}
\hat E_{\rm Rich}(0) = \hat c_0 = 
\sum_{k=1}^{m}\, y_k\, \prod_{i \neq k} \frac{\lambda_i }{\lambda_i - \lambda_k}\, , 
\end{align}
where we assumed that all the elements of $\bm \lambda$ are different. 

The error of the estimator is $O(m)$ only in the asymptotic limit $N \rightarrow \infty$. In other words $O(m)$ corresponds to the bias term in Eq.~\eqref{mse_zne}. In a real scenario, $N$ is finite, and the variance term in Eq.~\eqref{mse_zne} grows exponentially as we increase $m$.
This fact can be easily shown in the simplified case in which the noise scale factors are equally spaced, i.e., $\lambda_k = k\,\lambda_1$ where $k=1,2,\dots m$.
Substituting this assumption into Eq.~\eqref{richardson_analytic} we get:
\begin{equation}
    \label{eq:simplerichardsonintercept}
   \hat E_{\rm Rich}(0) = 
\sum_{k=1}^{m}\, y_k\, \prod_{i \neq k} \frac{i }{i - k}\ = \sum_{k=1}^m\, y_k\,(-1)^{k-1}\binom{m}{k}. 
\end{equation}
If we assume that each expectation value is sampled with the same statistical variance $\sigma^2$ as described in Eq.~\eqref{estim_dist}, since $\hat E_{\rm Rich}(0)$ is a linear combination of the measured expectation values $\{y_k\}$, its variance is given by:
\begin{align}
    \label{eq:var_E0}
    {\rm Var}(\hat E_{\rm Rich}(0))
    &= \sigma^2 \sum_{k=1}^m \binom{m}{k}^2 \nonumber\\
    &=\sigma^2 \left[ \binom{2m}{m} - 1\right] \xrightarrow{m \longrightarrow \infty} \sigma^2\  \frac{2^{2m}}{\sqrt{\pi m}},
\end{align}
where we used the Vandermonde's identity and, in the last step, the Stirling approximation.

The practical implication of Eq.~\eqref{eq:var_E0} is that the zero-nose limit predicted by the Richardson's estimator is characterized by a statistical uncertainty which scales exponentially with the number of data points.

\subsection{Poly-Exponential extrapolation}

The poly-exponential ansatz of degree $d$ is:
\begin{align}\label{poly_exp_ansatz}
 E_{\rm polyexp}^{(d)}(\lambda) = a \pm e^{z(\lambda)}, \; z(\lambda):= z_0 + z_1 \lambda+ \dots z_d \lambda^d.
\end{align}
where $a, z_0, z_1, \dots z_d$ are $d+2$ parameters.
From physical considerations, it is reasonable to assume that $E(\lambda)$ converges to a finite asymptotic value i.e.:
 \begin{align}\label{decay_hypothesis}
E(\lambda) \xrightarrow{\lambda \rightarrow \infty} a 
\quad \Longleftrightarrow \quad
z(\lambda) \xrightarrow{\lambda \rightarrow \infty} -\infty.
\end{align}
There are two important scenarios: (i) where $a$ is unknown and so a non-linear fit should be performed and (ii) where $a$ is deduced from asymptotic physical considerations. For example, if we know that in the limit of $\lambda \rightarrow \infty$ the state of the system is completely mixed or thermal, it is possible to fix the value of $a$ such that the poly-exponential ansatz \eqref{poly_exp_ansatz} is left with only $d+1$ unknown parameters: $z_0, z_1, \dots z_d$. 
If the asymptotic limit $a$ is known, we can apply the following procedure:

\begin{enumerate}
\item Evaluate $\{y_k'\} = \{ \log(|y_k-a| + \epsilon) \}$, representing the measurement results in a convenient logarithmic space with coordinates $(y_k', \lambda_k)$, with a small regularizing constant $\epsilon>0$. 
\item The model of Eq.~\eqref{poly_exp_ansatz} in the logarithmic space $(y_k', \lambda_k)$ reduces to the polynomial $z(\lambda)$.
\item Estimate the zero-noise limit in the logarithmic space $\hat z(0)=\hat z_0$ with a standard polynomial extrapolation. If necessary
different weights can be used for different scale factors, taking into account the non-linear propagation of statistical errors.
\item Convert back to the original space, obtaining the final estimator $\hat E(0)= a \pm e^{\hat z(0)}$.
\end{enumerate}

This allows us to map a non-linear regression problem into a polynomial fit that is linear with respect to the parameters and therefore much more stable. However, many reasonable alternative approaches exist like maximum likelihood optimization. Alternatively a Bayesian approach could be used, especially if we have prior information about the parameters of the model.

\subsection{Exponential extrapolation}

Exponential extrapolation is a particular case of the more general poly-exponential method. It corresponds to the model:
\begin{align}\label{exp_ansatz}
 E_{\rm exp}(\lambda) = E_{\rm polyexp}^{(d=1)}(\lambda) = a \pm e^{z_0 + z_1 \lambda}= a + b e^{-c \lambda},
\end{align}
where the set of real coefficients $a,b,c$ is a way of parametrizing the same ansatz, alternative but equivalent to $a, z_0, z_1$. This model was discussed in~\cite{endo2018practical} and is generalized by our extrapolation framework. In particular, increasing the order $d$, for example to $d=2$, and using the poly-exponential model  \eqref{poly_exp_ansatz} we can capture small deviations from the ideal exponential assumption,
possibly obtaining a more accurate zero-noise extrapolation.

\subsection{Benchmark comparisons of ZNE methods}

\begin{table}[tb]
%\centering
\begin{tabular}{l|l|p{12mm}|p{12mm}}
%\cline{1-3}
\bf Scaling & \bf Extrapolation & \bf Error \% \newline (dep.) & \bf Error \% \newline (amp. damp.) \\ %\cline{1-3}
\hline
\hline
none 	& unmitigated                 &    29.9\,$\pm$\,5.1      &    16.7\,$\pm$\,4.0 \\
\hline
circuit & linear $(d=1)$              &    14.6\,$\pm$\,4.6      &    5.40\,$\pm$\,2.3 \\
circuit & quadratic $(d=2)$           &    6.35\,$\pm$\,3.6      &    3.53\,$\pm$\,3.4 \\
circuit & Richardson $(d=3)$          &    17.6\,$\pm$\,11       &    \textcolor{red}{17.9\,$\pm$\,16}  \\
circuit & exponential $(a=0.25)$      &    2.73\,$\pm$\,1.9      &    2.06\,$\pm$\,1.6 \\
circuit & adapt. exp. $(a=0.25)$      &\bf 1.27\,$\pm$\,1.1      &    2.69\,$\pm$\,2.8 \\
\hline
at random & linear $(d=1)$            &    15.6\,$\pm$\,5.3      &    5.20\,$\pm$\,2.4 \\
at random & quadratic $(d=2)$         &    5.54\,$\pm$\,4.4      &    8.00\,$\pm$\,8.1 \\
at random & Richardson $(d=3)$        &  \textcolor{red}{30.0\,$\pm$\,24}     &   \textcolor{red}{24.0\,$\pm$\,18}  \\
at random  & exponential $(a=0.25)$   &    2.84\,$\pm$\,1.8      &\bf 0.95\,$\pm$\,1.0 \\
at random & adapt. exp. $(a=0.25)$    &    1.77\,$\pm$\,1.4      &    2.18\,$\pm$\,1.2 \\
\hline
from left & linear $(d=1)$            &    14.4\,$\pm$\,4.5      &    5.16\,$\pm$\,2.3 \\
from left & quadratic $(d=2)$         &    6.73\,$\pm$\,3.7      &    3.88\,$\pm$\,3.7 \\
from left & Richardson $(d=3)$        &    18.4\,$\pm$\,12       &    16.1\,$\pm$\,13  \\
from left & exponential $(a=0.25)$    &    3.17\,$\pm$\,2.1      &    2.19\,$\pm$\,2.0 \\
from left & adapt. exp. $(a=0.25)$    &    1.43\,$\pm$\,1.1      &    3.08\,$\pm$\,3.6 \\
\end{tabular}
\caption{Average of 20 different two-qubit randomized benchmarking circuits with mean depth 27. The percent mean absolute error from the exact value of 1 is reported for a depolarizing noise with $p=1\%$ and an amplitude damping channel with $\gamma=0.01$. For all non-adaptive methods we used $\lambda=\{1,1.5,2,2.5\}$. Adaptive extrapolation was iterated up to 4 scale factors.
All the results reported in this table are obtained with exact density matrix simulations. The best result for each noise model is highlighted with a bold font, while errors larger than the unmitigated one are red colored.}
\label{tab:benchmarks}
\end{table}
% TABLE DATA FROM https://github.com/unitaryfund/mitiq-examples/blob/master/random/RB_comparisons2.ipynb ("Second table with new scale factors")

Benchmarks comparing the performance of ZNE methods are given in Table~\ref{tab:benchmarks}. In all cases, besides for Richardson extrapolation, ZNE improves on the unmitigated noise value, however the performance varies significantly. Furthermore, one scaling or extrapolation method does not strictly dominate others. 

\begin{figure}[tb]
    \centering
    \includegraphics[width=\columnwidth]{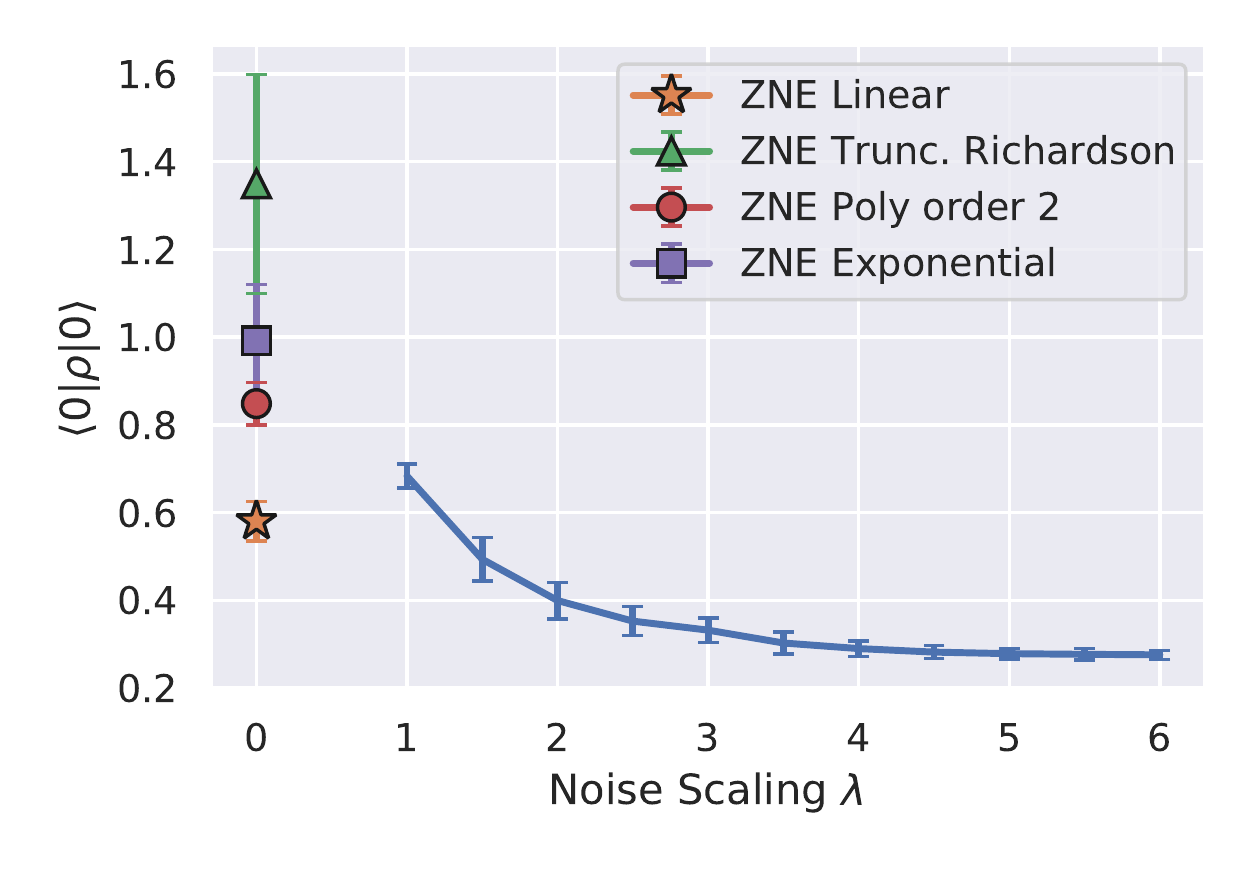}
    \caption{Comparison of extrapolation methods averaged over 50 two-qubit randomized benchmarking circuits executed on IBMQ's ``London'' five-qubit chip. The circuits had, on average, 97 single qubit gates and 17 two-qubit gates. The true zero-noise value is $\langle 0 | \rho | 0 \rangle = 1$ and different markers show extrapolated values from different fitting techniques.
    % WAS RELATED TO THE PREVIOUS PLOT:
    % All extrapolations are to zero noise, though they are separated on the plot for visualization.
    }
    \label{fig:ibmq}
\end{figure}

Different extrapolation methods are compared on IBMQ's London superconducting quantum processor in Fig.~\ref{fig:ibmq}. Here random gate folding scales the noise of 50 different two-qubit randomized benchmarking circuits. The ideal expectation value for all circuits is 1. The order 2 polynomial fit, and the exponential fit outperform Richardson extrapolation. In fact, Fig.~\ref{fig:ibmq} shows the expectation value for Richardson extrapolation when only the first 3 data points are considered. Instability in the Richardson extrapolation for more points, as described in Section~\ref{sec:richardson}, causes nonphysical results when applied to all the measured data. This is an example in which vanilla Richardson extrapolation is not sufficient to provide stable results.

% data available at https://github.com/unitaryfund/mitiq-examples/tree/master/ibmq/two-qubit-randomized-benchmarking

\section{Adaptive zero noise extrapolation}
\label{sec:adaptive}

In Section \ref{sec:non-adaptive}, we considered only non-adaptive extrapolation methods. However, in order to reduce the computational overhead, we can choose the scale factors and the number of samples in an adaptive way
as described in Algorithm~\ref{alg:adaptive}.

\begin{algorithm}[tb]
 \KwData{An initial set of $m$ noise scale factors $\bm \lambda = \{\lambda_1, \lambda_2, \dots \lambda_m \}$, with $\lambda_j \ge 1$, $m$ sample numbers $\bm N =(N_1, N_2, \dots N_m )$ and a maximum number of total samples  $N_{\text{max}}$.
 }
 \KwResult{A mitigated expectation value}
 \Begin{
 \tcc{Initialization}
 $\bm y \longleftarrow \emptyset$\;
 \For{$\lambda_j \in \bm \lambda$}{
    $y_j \longleftarrow ComputeExpectation(\lambda_j, N_j)$\;
    $Append\left(\bm y, y_j\right)$\;
    }
 \tcc{Adaptive loop}
 $N_{\text{used}} \longleftarrow 0$\;
 \While{$N_{\text{used}} < N_{\text{max}}$}{
  	$\bm \Gamma^*\longleftarrow BestFit(E_{\rm model}(\lambda; {\bm \Gamma}), (\bm \lambda, \bm y))$\;
    $\lambda_{\text{next}}\longleftarrow NewScale(\bm \Gamma^*, \bm \lambda, \bm y)$\;
    $N_{\text{next}}\longleftarrow NewNumSamples(\bm \Gamma^*, \bm \lambda, \bm y)$\;
    $y_{\text{next}} \longleftarrow ComputeExpectation(\lambda_{\text{next}}, N_{\text{next}})$\;
    $Append\left(\bm \lambda, \lambda_{\text{next}}\right)$\;
    $Append\left(\bm y, y_{\text{next}}\right)$\;
    $N_{\text{used}} \longleftarrow N_{\text{used}} + N_{\text{next}}$\;
 }
 \KwRet{$E_{\rm model}(0; {\bm \Gamma^*})$}\;
 }
 
 \caption{Generic adaptive extrapolation}
 \label{alg:adaptive}
\end{algorithm}

Differently from the non-adaptive case, in this adaptive procedure (Alg.~\ref{alg:adaptive}) the measured scale factors $\bm \lambda$ are not monotonically increasing. Indeed in the adaptive step, $\lambda_{\text{next}}$ can take any value (above or equal to $1$). In particular, $\lambda_{\text{next}}$ could also be equal to a previous scale factor $\lambda_j$, for some $j$. In this case, the additional measurement samples $N_{\text{next}}$ will improve the statistical estimation of $E(\lambda_j)$.

Now, we present an example of adaptive extrapolation which is based on the exponential ansatz $E_{\text{exp}}(\lambda)=a +b e^{-c \lambda}$ that we have already introduced in Eq.~\eqref{exp_ansatz}. We also assume that the asymptotic value $a$ is known. This implies that at least two scale factors should be measured to fit the parameters $b$ and $c$. We first consider this particular case and then we generalize the method to an a arbitrary number of scale factors, which will be chosen in an adaptive way.

\subsection{Exponential extrapolation with two scale factors}

We assume only two scale factors $\lambda_1$ and $\lambda_2$ (typically, $\lambda_1$ is $1$).
As discussed in Section~\ref{sec:non-adaptive}, we can estimate the corresponding expectation values, $E(\lambda_1)$ and $E(\lambda_2)$, with a statistical uncertainty of $\sigma_1^2=\sigma_0^2/N_1$ and $\sigma_2^2=\sigma_0^2/N_2$, respectively. Here, we are implicitly assuming that the single shot variance $\sigma_0^2$ is independent of $\lambda$, such that the estimation precision is only determined by number of  samples $N_1$ and $N_2$. The measurement process will produce two results $y_1$ and $y_2$, whose statistical distribution is given by Eq.~\eqref{estim_dist}.

Since the parameter $a$ is known, we can use the points $(\lambda_1, y_1)$ and $(\lambda_2, y_2)$ to estimate $b$ and $c$ of Eq.~\eqref{exp_ansatz}. The two estimators $\hat b$ and $\hat c$ can be determined by the unique ansatz interpolating
the two points, whose parameters are:
\begin{eqnarray}
    \hat c & = & \frac{1}{\lambda_2 - \lambda_1}\log\frac{ y_1 - a}{ y_2 - a}\, , \\
    \hat b & = & \left( y_1 - a\right)^{\frac{\lambda_2}{\lambda_2 - \lambda_1}}\left(y_2 - a\right)^{-\frac{\lambda_1}{\lambda_2 - \lambda_1}} \, .
    \label{eq:bformula}
\end{eqnarray}

The corresponding estimator for the zero-noise limit is $\hat E_{\text{exp}}(0)=a + \hat b$ where, since $a$ is known, the  error is only due to the statistical noise of $\hat b$.

This estimator depends on the empirical variables $y_1$, $y_2$, with statistical variances $\sigma_1^2=\sigma_0^2/N_1$ and $\sigma_2^2=\sigma_0^2/N_2$ respectively. Such measurement errors will propagate to the estimator $\hat b$. To leading order in $\sigma_1^2$ and $\sigma_2^2$,  we have: 
\begin{equation}\label{eq:mse_b}
  {\rm MSE}(\hat b)= \left(\frac{\partial \hat b}{\partial y_1}\right)^2 \sigma_{1}^2 + \left(\frac{\partial \hat b}{\partial y_2}\right)^2 \sigma_{2}^2\, .
\end{equation}

The explicit evaluation of Eq.~\eqref{eq:mse_b}, yields:

\begin{equation}\label{eq:mse_b_explicit}
  {\rm MSE}(\hat b)= \frac{\sigma_0^2}{(\lambda_2 - \lambda_1)^2}\left[
        \frac{\lambda_2^2\,e^{2c\lambda_1}}{N_1} + \frac{\lambda_1^2\,e^{2c\lambda_2}}{N_2}
    \right]\,.
\end{equation}
The previous equation shows that the error depends on the choice of the scale factors $\lambda_1$ and $\lambda_2$ but also on the associated measurement samples $N_1$ and $N_2$.

\subsubsection{Error minimization}

Let us first assume that we have at disposal only a total budget $N_{\text{max}}=N_1 + N_2$ of circuit evaluations and that $\lambda_1$ and $\lambda_2$ are fixed. 
Minimizing Eq.~\eqref{eq:mse_b_explicit}, with respect to $N_1$ and $N_2$, we get:
\begin{eqnarray}   
    N_1 & = & N_{\text{max}} \, \frac{\lambda_1}{\lambda_1 + \lambda_2\,e^{-c(\lambda_2 - \lambda_1)}} \, \nonumber \\
    N_2 & = & N_{\text{max}}\, \frac{\lambda_2\,e^{-c(\lambda_2 - \lambda_1)}}{\lambda_1 + \lambda_2\,e^{-c(\lambda_2 - \lambda_1)}} \,  \label{eq:samplingratio}
\end{eqnarray}
and the corresponding error becomes:

\begin{equation}
   {\rm MSE}(\hat b) = \sigma_0^2 \left[\frac{\lambda_2\,e^{c\lambda_1} + \lambda_1\,e^{c\lambda_2}}{\lambda_2 - \lambda_1}\right]^2 \,.
\end{equation}
This error can be further minimized with respect to the choice of the scale factors.
Since $\lambda_1$ is usually fixed to $1$, we optimize over $\lambda_2$, leading to the condition:
\begin{equation}
    e^{c(\lambda_2 - \lambda_1)}\left(c(\lambda_2 - \lambda_1) - 1\right) - 1 = 0 .
\end{equation}
We can solve the previous equation numerically, obtaining:
\begin{equation}
    \label{eq:numericalsoln}
    c(\lambda_2 - \lambda_1) = \alpha,
\end{equation}
where $\alpha\simeq 1.27846$ is a numerical constant.
For a fixed $\lambda_1$, the previous condition determines the optimal choice of the scale factor $\lambda_2$ which minimizes the zero-nose extrapolation error.
From a practical point of view, Eqs.~\eqref{eq:samplingratio} and~\eqref{eq:numericalsoln} can only be used if we have some prior knowledge about $c$. This motivates the following adaptive algorithm.

\begin{algorithm}[t]
 \KwData{An exponential model $E_{\text{exp}}(\lambda)=a + b e^{-c \lambda}$ with a known/estimated $a$. A maximum number of total samples $N_{\text{max}}$, a fixed number of samples per iteration $N_{\text{batch}}$ and
 a minimum scale factor $\lambda_1$ (typically equal to $1$).}
 \KwResult{A mitigated expectation value}
 \Begin{
  $c \longleftarrow 1 \text{;}$ \tcc{Initial guess}
  $\alpha  \longleftarrow 1.27846 \text{;}$ \tcc{Alpha in Eq.~\eqref{eq:numericalsoln}}
  ${\bf data} \longleftarrow \emptyset$\;
  $N_{\text{used}} \longleftarrow  0$ \;
 \While{$N_{\text{used}} < N_{\text{max}}$}{
    $\lambda_2 \longleftarrow \lambda_1 + \alpha / c$\;
    
    $
        N_1  \longleftarrow  N_\text{batch} \times \frac{c\,
        \lambda_1/\alpha}{c\, \lambda_1+\alpha -1} 
    $;
    \vspace{0.1 cm}
    
    $
        N_2  \longleftarrow   N_\text{batch}\times  \frac{(1 +c\,
        \lambda_1/\alpha) (\alpha -1)}{c\, \lambda_1+\alpha -1} 
    $\;
    $N_{\text{used}} \longleftarrow  N_{\text{used}}+ N_1 + N_2 $\;
    $y_1 \longleftarrow ComputeExpectation(\lambda_1, N_1)$\;
    $y_2 \longleftarrow ComputeExpectation(\lambda_2, N_2)$\;
    $Append\left({\bf data} , (\lambda_1, y_1)\right)$\;
    $Append\left({\bf data} , (\lambda_2, y_2)\right)$\;
    \tcc{New estimate of c}
 	$c\longleftarrow BestFit(E_{\rm exp}(\lambda; a,b,c),{\bf data} )$\;
 }
 \KwRet{$E_{\rm exp}(0; {a,b,c})$}\;
 }

\caption{Adaptive exponential extrapolation}
\label{alg:exp-adaptive}
\end{algorithm}

\subsection{An adaptive exponential extrapolation algorithm}

\begin{figure}[tb]
    \centering
    \includegraphics[width=\columnwidth]{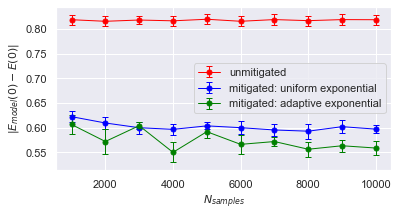}
    \caption{Comparison of adaptive and non-adaptive exponential zero noise extrapolation, given a fixed budget of samples. The adaptive method generally produces a more accurate extrapolation with less samples. On the other hand, in this example, the advantage of adaptivity is not particularly large. Likely, this is due to the fact that the scale factors used for the non-adaptive method are already quite good and not far from their optimal values. Data was generated by exact density matrix simulation of 5-qubit randomized benchmarking circuits of depth 10 under 5\% depolarizing noise and measured in the computational basis. Noise was scaled directly by access to the back-end simulator rather than with a folding method.}
    \label{fig:adaptive}
\end{figure}

Algorithm~\ref{alg:exp-adaptive} is an adaptive exponential algorithm based on the exponential ansatz $E_{\text{exp}}(\lambda)=a + b e^{-c \lambda}$, where $a$ is a known constant. Figure~\ref{fig:adaptive} shows a comparison of adaptive exponential extrapolation with non-adaptive exponential extrapolation. At almost all sample levels, adaptive extrapolation outperforms the non adaptive approach.

\section{Conclusion}

We make zero-noise extrapolation digital, developing the unitary folding framework to run error mitigation with instruction set level access. We then demonstrate improved performance through a set of non-adaptive and adaptive extrapolation methods. We emphasize that zero-noise extrapolation is in general an inference problem with many avenues for further optimization.

While ZNE has previously been benchmarked on randomized benchmarking circuits or VQE, we give benchmarks of ZNE on MAXCUT problems solved with QAOA. This allows us to smoothly benchmark the performance of ZNE on larger variational quantum circuits then have been considered previously.

We also consider specialization of zero-noise extrapolation to different noise models, using calibration noise as an example. With more sophisticated multi-parameter noise models (such as a combination of calibration noise and amplitude dampening), it is likely that multi-dimensional noise extrapolation~\cite{otten2019recovering} will be of interest. 

This work is a first step towards viewing zero-noise extrapolation as an inference problem and has opportunities for extension. Priors or constraints from observable, noise or circuit structure could be included. Data could be gathered from similar executions over time so that inference includes a historical database of previous computations. 

Error-mitigation is likely to remain a critical toolkit for the NISQ-era quantum programmer. Improving and benchmarking these techniques will likewise remain an important task.

\section*{Acknowledgment}
We thank Jonathan Dubois and Lorenza Viola for helpful discussions on the calibration noise model and Nathan Shammah for feedback on the manuscript. This material is based upon work supported by the U.S. Department of Energy, Office of Science, Office of Advanced Scientific Computing Research, Accelerated Research in Quantum Computing under Award Number DE-SC0020266 and by IBM under Sponsored Research Agreement No. W1975810. We thank IBM for providing access to their quantum computers. The views expressed in this paper are those of the authors and do not reflect those of IBM.

\bibliographystyle{ieeetr}
\bibliography{references.bib}

\end{document}